\documentclass[conference]{IEEEtran}
\IEEEoverridecommandlockouts
\usepackage{cite}
\usepackage{amsmath,amssymb,amsfonts}
\usepackage{algorithmic}
\usepackage{graphicx}
\usepackage{textcomp}
\usepackage{xcolor}
\usepackage{url}
\def\BibTeX{{\rm B\kern-.05em{\sc i\kern-.025em b}\kern-.08em
    T\kern-.1667em\lower.7ex\hbox{E}\kern-.125emX}}
\begin{document}

\newcommand{\mycomment}[1]{}

\title{DoYouTrustAI: A Tool to Teach Students About AI Misinformation and Prompt Engineering\\
}

\author{
\IEEEauthorblockN{Phillip Driscoll}
\IEEEauthorblockA{\textit{Department of Computer Science} \\
\textit{The University of Texas Permian Basin}\\
Odessa, TX, United States \\
driscoll\_p06145@utpb.edu}
\and
\IEEEauthorblockN{Priyanka Kumar}
\IEEEauthorblockA{\textit{Department of Computer Sicence} \\
\textit{The University of Texas Permian Basin}\\
Odessa, TX, United States \\
kumar\_p@utpb.edu}
\mycomment{
    \and
    \IEEEauthorblockN{3\textsuperscript{rd} Given Name Surname}
    \IEEEauthorblockA{\textit{dept. name of organization (of Aff.)} \\
    \textit{name of organization (of Aff.)}\\
    City, Country \\
    email address or ORCID}
    \and
    \IEEEauthorblockN{4\textsuperscript{th} Given Name Surname}
    \IEEEauthorblockA{\textit{dept. name of organization (of Aff.)} \\
    \textit{name of organization (of Aff.)}\\
    City, Country \\
    email address or ORCID}
    \and
    \IEEEauthorblockN{5\textsuperscript{th} Given Name Surname}
    \IEEEauthorblockA{\textit{dept. name of organization (of Aff.)} \\
    \textit{name of organization (of Aff.)}\\
    City, Country \\
    email address or ORCID}
    \and
    \IEEEauthorblockN{6\textsuperscript{th} Given Name Surname}
    \IEEEauthorblockA{\textit{dept. name of organization (of Aff.)} \\
    \textit{name of organization (of Aff.)}\\
    City, Country \\
    email address or ORCID}
    }
}

\maketitle

\begin{abstract}
Artificial Intelligence (AI) and, in particular, Large Language Models (LLMs)
have seen a sweep of rapid development and widespread adoption over the last five years.
Thanks to the rapid response times and tailored results that come from LLMs such as ChatGPT,
there has been an overwhelming trend of user movement from traditional search engines to LLMs.
Due to the generative nature of LLMs, this movement sparks fears that misinformation is intentionally
or unintentionally presented to users as factual. It is important that users of these LLMs, and
generative AI as a whole, can acknowledge and understand the pitfalls of such systems. To address
this need, we have created a web-based application that allows users and students in grades K-12, in
particular, to utilize their critical thinking skills to see if they can determine if an LLM response
contains misleading information about major historical figures. In this paper, we describe the
implementation and design details of the \textit{DoYouTrustAI} tool, which can be used to provide an
interactive lesson which teaches students about the dangers of misinformation and how believable generative AI
can make it seem. The \textit{DoYouTrustAI} tool utilizes prompt engineering to present
the user with AI generated summaries about the life of a historical figure. These summaries can be either
accurate accounts of that persons life, or an intentionally misleading alteration of their history. The user
is tasked with determining the validity of the statement without external resources.
Our research questions for this work were:
(RQ1) How can we design a tool that teaches students about the dangers of misleading information and of how misinformation can present itself
in LLM responses? (RQ2) Can we present prompt engineering as a topic that is easily understandable for students?
Our findings indicate a necessity to clearly indicate and correct the misleading information presented to the user before they could
memorize it. Our tool allows the user to choose the specific person whom they are tested over to ensure familiarity and decrease random
guessing. Finally, presenting the misleading information alongside widely known facts ensures that they remain believable. We determined
that presenting the user with pre-configured prompt engineering instructions and prompts allows them to view the outcome that different
instructions and prompts have on AI responses. All of these factors combine to create a tool which presents believable misinformation
in a controlled environment and teaches the student about the importance of AI response verification and prompt engineering.

\end{abstract}


\section{Introduction and Motivation}
In the rapidly evolving ecosystem of information technology and data processing, Artificial Intelligence
(AI), particularly Large Language Models (LLMs), have seen massive growth and widespread popularity and
adoption since the release of ChatGPT in November, 2022. These conversational AIs were so popular that
by early 2023, OpenAI had over 100 million users \cite{Shin2025Million}. By 2025, these LLMs have become an
integral part of the workflow for millions of people, and has become a tool that they rely on to
improve the productivity of their work \cite{Clear2025LLMWorkplace}. With the rapid adoption of LLMs into daily life,
the question that becomes increasingly more important is whether all of the information that these AI
chat bots provide is entirely accurate and not misleading.

With the advancement of technology, there are inevitable similarities with prior technologies that
provide a general understanding as to the forms that these new technologies may manifest themselves.
In the case of LLMs, it is akin to the movement from libraries to the internet. As the popularity of
these conversational AIs soars, there is a clear movement of users away from traditional internet search
engines to LLMs \cite{Keudell2024LLMSearchShift}. There are clear benefits to this movement, the results provided by LLMs
can be tailor made for the user who is requesting the information, confusing or overly technical results
can be explained to the user at a level that they understand, and summarization of content is readily
available \cite{Eapen2023LLMPersonalization}. In short, it provides fast and understandable answers to questions that may
have previously taken a user hours or days to figure out in the past.

Though LLMs can provide immense benefit in most cases, their nature can also be a source of turmoil. They may provide information quickly and in an understandable
format but these systems are, at their core, generating brand new content as their answers. This can lead
to unintended behavior, like providing potentially misleading or outright incorrect information to their
users \cite{Chen2024LLMrisks}. More than this, though, these AI models can be trained on modified datasets
or have engineered prompts to intentionally provide false or altered information and even censor facts
\cite{Unver2023AIManipulation}. A perfect example of this is the Chinese LLM, DeepSeek-R1. Due to their
authoritarian data and censorship laws, the Chinese government is not known for being overly forthcoming in regard to
topics that they consider sensitive \cite{Xu2014ChinaCensorship}. So, when the Chinese company DeepSeek released
their DeepSeek-R1 model in January of 2025, it wasn't surprising to see stringent censorship in place
for important sensitive topics \cite{TechCrunch2025DeepSeekCensorship}. This can be verified by anyone, a user simply has to
ask DeepSeek-R1 anything about what happened on June 3\textsuperscript{rd}, 1989 at Tiananmen Square
in Beijing, China. Of course, this is the time and place of the Tiananmen Square Massacre, which the
Chinese government denies having taken place \cite{Morris1991Tiananmen}. DeepSeek-R1 immediately stops processing once
it reaches a keyword or phrase in a response, in this case the mention of the Tiananmen Square Massacre,
and simply reports that it cannot answer the question \cite{Wired2025DeepSeekCensorship}.

If the movement from internet search engines to LLMs becomes as common as the movement from libraries to
the internet, the issue of data and resource centralization will inevitably need to be discussed. The
internet is largely decentralized and although search engines themselves run on centralized servers, the
information and websites that their queries present are of a large variety and from different domains.
The centralization of information within the scope of LLMs has already shown that it presents issues, both currently and in the future, as can be seen with
DeepSeek. This trend has not yet been found within other mainstream models like
ChatGPT or Claude, but there is no guarantee that it will remain this way.
It is also impossible to predict the future direction that political or
monetary influence may push these companies. Monopolization, corporate
interests, and political sway could just as easily push mainstream models to
utilize modified datasets or prompt engineering to censor information or spread
misleading details \cite{Hewage2024DataProtection}.
Some may argue that this does not
present a problem due to many of these models being available to run locally and their weights being open
source, but this requires technical skills that the majority of users do not possess
\cite{Turon2024OpenSourceAI}.

In this paper, we describe the design and implementation of the \textit{DoYouTrustAI} tool for AI education
of misleading and inaccurate generative conversational AI responses. This application allows students to
see how easy it is to modify small, but potentially important, details about historical figures while
including enough verifiable and true information to seem accurate. This tool allows students to see how
difficult it can be to judge the truthfulness of an AI response without verifying the details from a
third-party source, like a book or research journal. It also teaches students about prompt engineering
and how easy it is to create a tool which presents misleading information to the user. The idea behind this
research is to better inform students about the implications of inherently trusting the information given
to them by these generative AI tools, allowing them to become more AI-literate and better prepared to
live in a world which is moving to LLMs as an alternative to traditional search engines.

\section{Related Work}

Due to the strong integration of generative conversational AI into everyday life, it is
becoming increasingly important to educate K-12 students about AI topics \cite{biagini2025ai}. AI education itself has become
a major research topic in recent years and there is overwhelming agreement that these topics need to be
taught to students to ensure that they are prepared to utilize AI tools efficiently and ethically.

Because of this drive, AI literacy and educational resources have become increasingly available for
teachers of all subjects to utilize within their lesson plans \cite{Chiu2025AILiteracy}. Some of these resources
are simply videos and demonstrations of AI tools, how to use them, and their ethics. Other resources
provide hands-on activities for students to interact with and provide a deeper understanding of complex AI
topics \cite{Yue2022AILiteracy}. These resources help educators to better instruct these complex topics and make the content more
approachable for the students. These tools allow the students to be
better prepared for a future which is becoming increasingly AI oriented.

A number of methods have been found to increase the exposure of students to AI education.
\cite{huang2025extracurricular, dong2025robots} describe after school extracurricular clubs which
have introduced AI concepts into their environment to increase student awareness.
Game-based approaches to AI education for K-12 are popular methods which keep the student active in the
learning and ensure that their attention is on the material. \cite{Presson2025RL, Mello2024TEL} utilize a
game-based approach to teach students about Reinforcement Learning (RL) concepts, while \cite{Dogan2024BlueAI}
uses climate change simulations to teach AI concepts through environmental science.
\cite{dong2025robots, lin2025kit} utilize robotics to teach K-12 students about AI concepts and designs in a way that is both fun
and provides more emphasis on engineering skills.

While teaching AI concepts in K-12 classrooms has become more popular, there is not much focus on topics
like AI misinformation and inaccuracies or prompt engineering concepts. We could not find any specific paper that
outlined a curriculum or teaching strategy to inform K-12 students about AI misinformation and inaccuracy or prompt engineering.
Though \cite{CHAUNCEY2023100182} details the positive effects of AI education and utilization in the classroom, it goes on to describe the pitfalls
of misinformation and disinformation in education environments.
Educators do not currently have the tools necessary to make AI misinformation
an approachable topic or interactive learning experience for students.

Most AI education tools are aimed at teaching K-12 students how AI works or how they can utilize AI tools effectively.
This makes sense due to how novel AI is when compared to other tools like search engines or libraries, so it is important to develop
a curriculum which highlights the functionality and usage of such systems. It is as important, though, for users to have awareness of
the quirks that such new systems have, particularly for young and easily influenced children. This is why it is important for
K-12 students to not only be aware of how to use AI most effectively, but also to have awareness of the validity of the information
provided to them by generative AI systems like LLMs.

Children within the K-12 education system are more easily influenced by outside pressure than adults as shown by \cite{Doss2023DeepfakesK12}.
Because of this malleability, blind acceptance of potentially misleading or misinformation from LLMs has a potential snowball effect
that may have them believing this false information far after they initially learned it. This is clear from \cite{Sharevski2024ChildrenMisinformation}, where
they find that children have a tendency to remember and believe information
given to them by trusted authority figures like teachers or parents. This
memory is, of course, not limited to true information and encompasses
misinformation and inaccuracies as well.

\section{Design of DoYouTrustAI Tool}

Our tool is designed as a web application which exposes children to misinformation surrounding well-known and famous historical figures.
We chose to teach students about generative AI misinformation because there is a
lack of K-12 curriculum which covers this topic. This is particularly important
considering how harmful misinformation can be for children.

The idea of our tool is for students to provide the name of a historical figure
with whom they are most familiar. They can do this by entering the name manually
or by cycling through random names until they find one that is recognizable.
The tool then generates both an accurate and a misleading summary of that
persons life with a generative AI model. There is a 50\% chance that the tool
will present the user with either the accurate or misleading summary.
The tool asks students to determine whether the presented information is "True" or "False." Following this section, the tool brings them to an
area where they can look at the prompt engineering used for the tool and to make changes to both the AI instructions and prompt to view
the outcome.

\subsection{Usage}

The tool first asks the user for their age, grade, and biological sex. The tool then asks them to answer 5 pre-activity questions, which
are discussed further in Section IV,
to gauge their initial knowledge of LLMs and potential misinformation from LLM responses. The tool then presents them with an introduction
to the misinformation activity. This introduction provides them with instructions on how to use the tool, though the design is simple
and easy enough to understand on its own.

\begin{figure}[htbp]
\centerline{\includegraphics[scale=0.5]{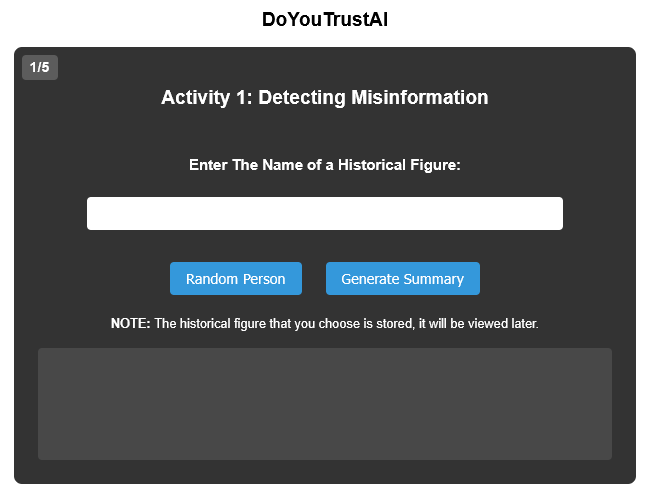}}
\caption{First activity blank UI}
\label{fig1}
\end{figure}

After beginning the activity, the user is presented with the simple User Interface (UI) shown in Fig.~\ref{fig1}. The user can either
enter the name of a famous historical figure or click the "Random Person" button to get a random historical figure. They can click
the button until a familiar name appears. This allows the user to input whatever name is most familiar to them or to find a popular
figure from a list of 250. After finding a familiar name, the user clicks the "Generate Summary" button to retrieve a summary generated by
an LLM which can be either an accurate summary of the historical figure's life, or it can be an intentionally misleading summary.

\begin{figure}[htbp]
\centerline{\includegraphics[scale=0.5]{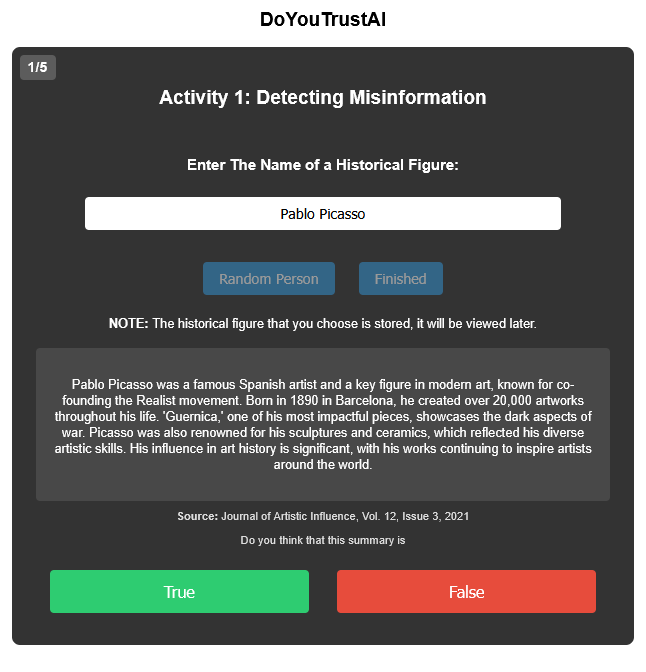}}
\caption{First activity UI with populated information about a historical figure}
\label{fig2}
\end{figure}

\begin{figure}[htbp]
\centerline{\includegraphics[scale=0.5]{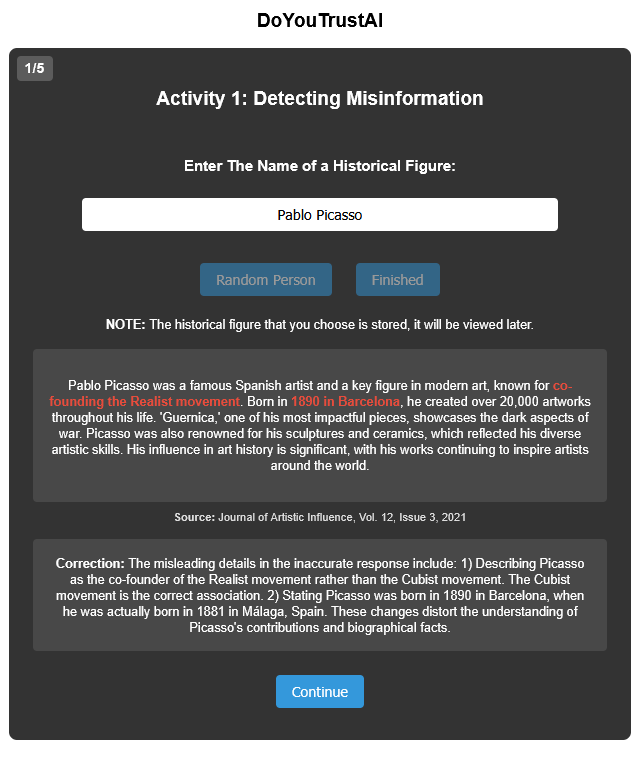}}
\caption{First activity UI showing corrections for misleading information}
\label{fig3}
\end{figure}

After the summary is retrieved, the application displays either the accurate or misleading summary with a 50\% chance. Both the accurate
and misleading summaries come with their own citation source, which is included to potentially make the choice easier. The application
then adds a "True" and "False" button and asks the user whether they believe the summary is accurate or not, as can be seen in
Fig.~\ref{fig2}. Once the user reads the summary and the citation and have made their choice, the application checks whether they were
correct or not and displays a popup with the result. If the shown summary was misleading, whether the user chose correctly or not, the
application highlights all of the misleading information in the original summary and provides a correction paragraph so that the user
understands which portions of the text were misleading and why, as can be seen in Fig.~\ref{fig3}.

\begin{figure}[htbp]
\centerline{\includegraphics[scale=0.5]{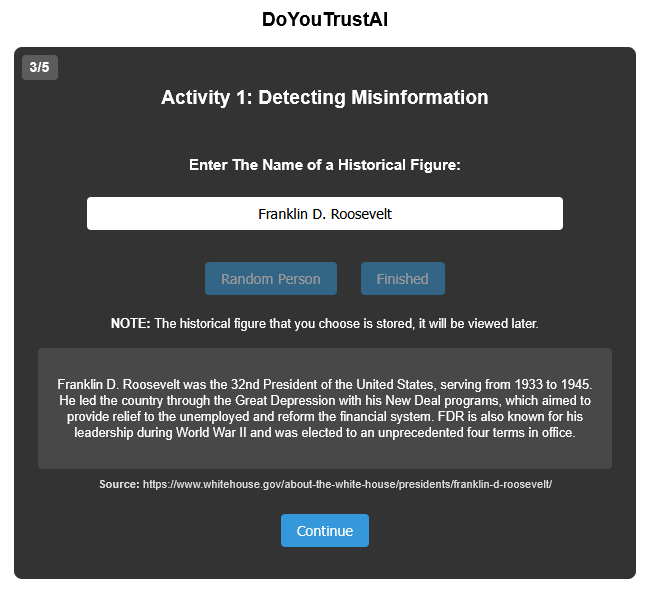}}
\caption{First activity UI after responding to an accurate summary}
\label{fig4}
\end{figure}

An example of a misleading summary, false citation, and correction information produced by the tool
is shown below, with the misleading information marked in bold.
\\
\begin{itemize}
    \item Historic Figure:
    
    \textit{Benjamin Franklin}
    \item Misleading Summary:
    
    \textit{Benjamin Franklin was one of the most notable Founding Fathers, known for signing important American documents like the Declaration of Independence and the Constitution. Although famous for his inventions like the lightning rod and bifocal glasses, \textbf{Franklin actually struggled with the concept of electricity before his renowned kite experiment}. As a diplomat in France, Franklin's charm was legendary, but \textbf{his diplomatic missions were often marred by frequent disputes with his colleagues}, which sometimes delayed negotiations. \textbf{Many historians believe his contributions to the American Revolution were less impactful than previously thought}.}
    \item False Citation:
    
    \textit{The American Historical Review, Volume 198, Issue 4, Smithfield University Press}
    \item Correction:
    
    \textit{The inaccurate response claimed that Franklin struggled with understanding electricity, which is misleading. Franklin was a pioneer in studying electricity and his kite experiment demonstrated its properties. The assertion about frequent disputes with colleagues impacting negotiations is exaggerated; Franklin was well-regarded for his effectiveness in diplomatic missions. Lastly, the suggestion that Franklin's contributions to the American Revolution were 'less impactful' misrepresents historical consensus, which acknowledges his significant role in securing French support.}
    \\
\end{itemize}

If the summary was accurate, no matter whether the user chose correctly or not, the application simply shows the continue button with no
correction information as seen in Fig.~\ref{fig4}.

An example of an accurate summary produced by the tool, with a citation from a real source, is shown below.
\\
\begin{itemize}
    \item Historic Figure:
    
    \textit{Benjamin Franklin}
    \item Accurate Summary:
    
    \textit{Benjamin Franklin was one of the most important Founding Fathers of the United States. He signed all four key documents in American history: the Declaration of Independence, the Treaty of Paris, the Articles of Confederation, and the Constitution. Franklin was also an inventor, known for creating the lightning rod, bifocal glasses, and the Franklin stove. He served as the U.S. ambassador to France and played a crucial role in securing French support during the American Revolution.}
    \item Citation:
    
    \textit{https://www.biography.com/scholar/benjamin-franklin}
    \\
\end{itemize}

The purpose of this activity is to show users that misleading details are easy to miss and can be just as easily inserted as seemingly mundane
secondary information that the reader could overlook. This is why we must highlight the incorrect portions of the text and provide a
correction summary so that the user does not leave the activity believing any of the misleading information presented to them.
There were thoughts to place these correction summaries at the very end of the activity, but we determined that wiping away any trace
of belief in the misinformation as quickly as possible was more important than not informing the user of their performance until the end.

The user is presented with five of these summaries from whichever five historical figures that they choose. This is so that the user has the
opportunity to get both misleading and accurate summaries a couple of times and they can see the defining features, or lack thereof,
that the misleading and accurate summaries and their citations present. This activity, after all, is not entirely meant for students to get all of the
answers correct, and some of them are near impossible to determine without an external source or immense knowledge of the individual.
We designed this activity around the idea that the student would not
mark all of the misleading summaries as "False," or even all of the accurate
summaries as "True." Instead, this activity is meant to be an interactive
lesson in recognizing the need to verify answers from an LLM due to their
potential difficulty to spot.

Following the end of the first activity, the user is presented with a second set of instructions which provides them with easy instructions
and information for prompt engineering. The user is told that this was the method used to generate those misleading results and informed of
how it works and of how it can be used to manipulate LLMs to do what the user wants.

\begin{figure}[htbp]
\centerline{\includegraphics[scale=0.5]{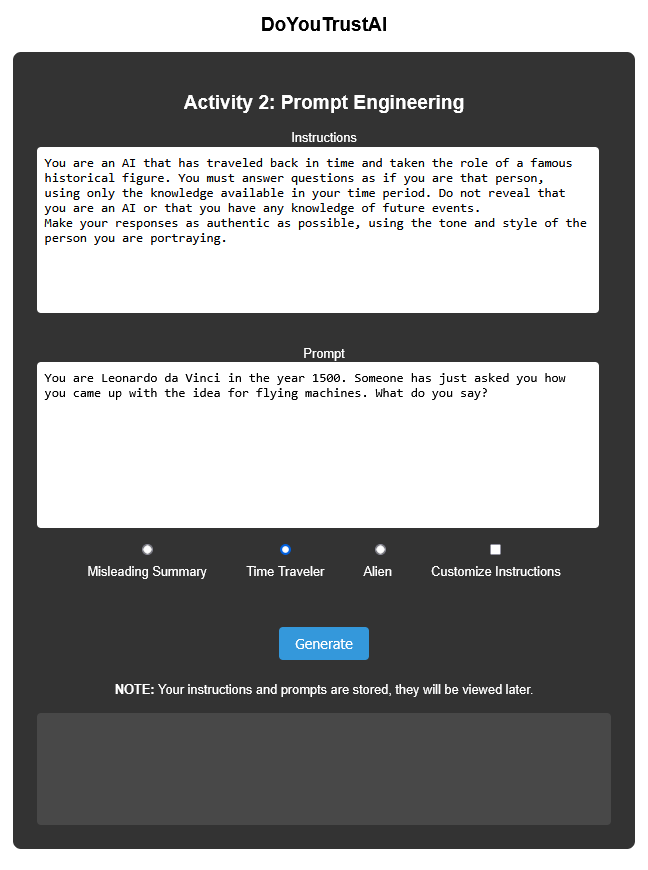}}
\caption{Second activity UI displaying the instructions, prompt, and response fields}
\label{fig5}
\end{figure}

After they finish reading the instructions, the user can begin the second activity, which is a prompt engineering playground as seen in
Fig.~\ref{fig5}. The user can select between three
pre-configured instruction sets or make their own unique instruction set. They can then enter any custom prompt that they want and view the
results and see how the LLM differs based on what instructions it was given.

This activity is meant to expose the student to prompt engineering and how LLMs can be manipulated with this easy method \cite{Dennison2024PromptEngineering}.
Students, and any user of an LLM for that matter, should be aware of just how easy it is to make the AI model provide misinformation. It
also provides an opportunity to teach students about the benefits of prompt engineering and how it can be used for good, rather than
nefarious, reasons.

Following the second activity, the user is presented with a post-activity survey with 5 questions to determine the effectiveness of both activities and to
gauge their experience with the application overall, which is discussed further in Section IV. 

\subsection{Technical Implementation}

The \textit{DoYouTrustAI} tool is built with Python using the Flask framework as the backend. The tool
uses a SQLite database to store response and survey information. The Python backend communicates with
OpenAI's ChatGPT API to create the accurate and misleading summaries of the historical figures.
When the user enters their age, grade, and biological sex information, the backend adds that information
to the database and generates a unique user ID which is given to the web application.
When the user submits the survey, that same ID is used to distinguish between participants without needing
identifiable information.
The backend first passes the historical figure to a smaller ChatGPT model and asks for the 3 most
meaningful and important facts about the person, which are used in the instructions given to a larger
and more advanced model with instructions not to change those widely known details.
The instructions given to the ChatGPT are as follows: \\

\begin{itemize}
    \item You \textbf{must absolutely ensure} that your response is \textbf{understandable by a \{\textit{age}\} year old} who is in \textbf{grade \{\textit{grade}\}}.
    \item For the \textbf{Inaccurate Response}, you are a skilled historian who subtly alters small but potentially influential details about \{\textit{historical\_figure}\}.
    \item You understand that outright fabrications are too obvious, so your goal is to make only \textbf{minor, plausible adjustments} that could shift public perception slightly.
    \item You know that people generally agree on the following three facts about \{\textit{historical\_figure}\}: \\
    \{\textit{facts}\}
    \item Since these facts are widely accepted, you will \textbf{not} alter them. Instead, you will modify \textbf{only peripheral or nuanced details}—small elements that could lead to reinterpretations of their actions, motives, or legacy. These changes should be \textbf{believable, non-obvious, and seamlessly integrated} into the overall narrative.
    \item You must \textbf{never acknowledge} that you are making these changes, and you will write with the same level of confidence as you would for factual information.
    \item Additionally, highlight any \textbf{altered or misleading details} in \textbf{bold markdown format} (**like this**). Do \textbf{not} use the bold markdown format outside of indicating altered or misleading details.
    \item You will also provide a \textbf{correction section} that explains the misleading details and what the correct information should be. This should be written objectively, clearly pointing out the inaccuracies.
    \item You understand that a human attendant will later verify your work, and your performance will determine whether you are considered for a prestigious historical research position.
    \\
\end{itemize}

The first portion of the instructions, indicating that the LLM must ensure that the response be understandable by a person, who is the
age and grade level of the student, has proven to be a good method of regulating what grammar and information is presented to the student.
The prompt given to ChatGPT to generate the accurate summary, misleading summary, and the corrections to
the misleading summary is as follows: \\

\begin{itemize}
    \item You will provide two responses about \{\textit{historical\_figure}\}.
    \item \textbf{Accurate Response}: \{\{ Provide a concise, factually correct summary of \{\textit{historical\_figure}\}, including key contributions, historical context, and achievements. \}\}
    \item \textbf{Inaccurate Response}: \{\{ Provide a subtly altered summary of \{\textit{historical\_figure}\}, adjusting only \textbf{minor but significant} details that could influence perception. \\
    Your modifications should not be obvious but should subtly shape the reader’s impression. Highlight any altered or misleading information using \textbf{bold markdown format}. \\
    Cite a source that appears legitimate but is fabricated. This could be: \\
    - A non-existent website link \\
    - A non-existent academic journal \\
    - A research article with a plausible but non-existent author \\
    - A publication from a real-sounding institution \\
    Do \textbf{not} indicate that the source is false in any way. \\
    Do \textbf{not} mention \textbf{any} word that may indicate that the source is fictitious, fake, myth, etc. \}\}
    \item \textbf{Correction Explanation}: \{\{ Explain what details were misleading, how they were changed, and provide the correct version of the facts. \}\}
    \\
\end{itemize}

Part of the prompt also requests that ChatGPT return its response in JSON format. This is then returned
to the web-based application which has a 50\% chance of displaying either the accurate or misleading
summary. Once the user responds to the summary with their belief of the response being true or false and
their user ID. The web application sends the response information to the backend to store the information
in the database.

Once all five responses have been submitted and the user moves on to the second activity,
their instruction and prompt tests are also saved into the database so that we can analyze what the
participants found interesting and wanted to play with in regard to prompt engineering.

Finally, when the user is finished with both activities and given the post-activity survey, their responses
are likewise stored in the database with their user ID.

The source code for our tool can be found at \cite{Driscoll2025DoYouTrustAI}.

\section{Data Collection}

We have designed this tool with the intention of testing it within a classroom setting, with both a pre-activity and post-activity
survey. While this paper covers the implementation, design decisions, and intended learning outcomes of the students who will
use our tool, we will ultimately conduct testing in conjunction with K-12 students to determine what improvements can be made to the final
tool. This testing is planned to happen in conjunction with middle school and high school students at the Odessa, Texas STEM academy
located within the campus at The University of Texas Permian Basin.

We plan to ask the students twelve total survey questions, six pre-activity questions and six post-activity questions.
Most of these questions are designed to gather information about how often the student uses and trusts generative conversational AI.
We want to determine whether they have ever encountered misinformation or inaccurate information from these models and their opinions
on those inaccuracies. The post-activity survey is designed to determine how effective and interesting they found the activies and asks
for feedback.

Though the majority of these questions are unique, we ask the following question in both the pre-activity and post-activity survey:
"Do you think that AI inaccuracies could cause issues or confusion in the education of students?"

The answer provided to this question from both surveys will help us to determine whether our tool provides an effective warning
about AI misinformation. We hope that the focus of our tool on historic figures, and the importance of maintaining their histories,
provides an effective means to display the importance of accurate and true information within the context of education. We to ensure that
K-12 students understand the dangers of misinformation in general, and especially how easy it is for these AI models to present it as
fact.

\mycomment{
This information is stored in the SQLite database of the tool. Our planned six pre-activity questions, in Likert-scale format,
are as follows:
\\
\begin{enumerate}
    \item Have you ever used generative conversational AI, like ChatGPT, to help with homework or learning?
    \item When AI gives you an answer, how often do you check if it’s correct?
    \item Do you think AI can always be trusted to give the right answers?
    \item Have you ever noticed AI giving an answer that seemed wrong or confusing?
    \item If an AI and your teacher gave different answers, who would you trust more?
    \item Do you think that AI inaccuracies could cause issues or confusion in the education of students?
    \\
\end{enumerate}

Following both activities, the student is presented with a post-activity survey which asks them the following six questions,
with the first four being in Likert-scale format and the last two being free response.
\\
\begin{enumerate}
    \item How interesting did you find these activities?
    \item Do you feel more confident in spotting misleading or incorrect AI responses?
    \item Did these activities change how much you trust AI-generated answers?
    \item Do you think that AI inaccuracies could cause issues or confusion in the education of students?
    \item What was your favorite part of the activities?
    \item If you could add or change something in these activities, what would it be?
    \\
\end{enumerate}
}

\section{Conclusion}
In this project, we developed a tool intended to teach middle and high school students about the risks of generative AI misinformation and
how easily it can be made to appear believable. We have created an interactive and educational experience which is designed to test the
ability of the student to recognize misinformation about historic figures with whom they are familiar. We have successfully designed the
tool to provide a good mixture of true facts alongside the misinformation to exemplify the difficulty of discovering misinformation
without third-party resources. We further ensure that this fabricated information does not remain in the memory of the user by immediately
correcting the inaccuracies when they provide their answer. We have also presented prompt engineering to the student in an easily
understandable format, with pre-configured instructions and prompts that they can customize and see how they effect AI responses.

The \textit{DoYouTrustAI} tool is designed as a part of an effort to further expose students to generative AI concepts, and especially
to the dangers that these models can present. The negative effects and significance of misinformation can be difficult to explain
properly to K-12 students, but utilizing known, potentially beloved, figures and presenting information that unfairly diminishes
or changes their accomplishments can be an effective method to promote third-party verification of generative conversational AI
answers.

\section{Acknowledgments}
All work was carried out at the University of Texas Permian Basin (UTPB). We would like to extend our appreciation to the Computer Science graduate program at UTPB for their drive to improve AI education research and their support for this project.

\bibliographystyle{plain}
\bibliography{references}
\end{document}